\documentclass[12pt]{article}
\usepackage{epsfig,amssymb,amsmath,psfrag}

\catcode`\@=11
\DeclareMathAlphabet{\mathpzc}{OT1}{pzc}{m}{it}
\newcommand{\insertfig}[2]{\mbox{\epsfxsize=#1cm \epsfbox{#2.eps}}}
\font\cmss=cmss12 
\def\1{\hbox{{1}\kern-.25em\hbox{l}}}
\def\bfZ{\relax{\hbox{\cmss Z\kern-.4em Z}}}
\def \be  {\begin{equation}}
\def \ee  {\end{equation}}
\def \ba  {\begin{eqnarray}}
\def \ea  {\end{eqnarray}}
\def \baa {\begin{eqnarray*}}
\def \eaa {\end{eqnarray*}}
\def \bb  {\begin {thebibliography} }
\def \eb  {\end{thebibliography}}
\def \lab #1 {\label{#1}}

\newcommand\re[1]{(\ref{#1})}

\def \matrix #1 {\left(\begin{array}{cc} #1 \end{array}\right)}

\newcommand{\as}{\ifmmode\alpha_{\rm s}\else{$\alpha_{\rm s}$}\fi}
\newcommand{\asbar}{\ifmmode\bar{\alpha}_{\rm s}\else{$\bar{\alpha}_{\rm s}$}\fi}

\newcommand{\ft}[2]{{\textstyle\frac{#1}{#2}}}

\newcommand{\pa}{\partial}

\newcommand{\cN}{{\cal N}}

\renewcommand{\a}\alpha

\newcommand{\bt}[1]{{\bar t}}

\font\cmss=cmss12 
\def\inbar{\,\vrule height1.5ex width.4pt depth0pt}
\def\IC{\relax\hbox{$\inbar\kern-.3em{\rm C}$}}
\def\IZ{\relax{\hbox{\cmss Z\kern-.4em Z}}}
\def\IR{{\hbox{{\rm I}\kern-.2em\hbox{\rm R}}}}

\def\IP{{\hbox{{\rm I}\kern-.2em\hbox{\rm P}}}}
\def\II{\hbox{{1}\kern-.25em\hbox{l}}}

\def\numberbysection{\@addtoreset{equation}{section}
                     \def\theequation{\thesection.\arabic{equation}}}
\numberbysection

\newbox\lett\newdimen\lheight\newdimen\lwidth
\def\ontop#1#2{\setbox\lett=\hbox{#2}\lheight\ht\lett
\multiply\lheight by 12 \divide\lheight by 10\relax%
\lwidth\wd\lett \multiply\lwidth by 8 \divide\lwidth by 10\relax #2\kern-\lwidth%
\raise\lheight\hbox{{$\scriptstyle #1$}}\kern.1ex}

\def\XXint#1#2#3{{\setbox0=\hbox{$#1{#2#3}{\int}$}
     \vcenter{\hbox{$#2#3$}}\kern-.5\wd0}}

\begin{document}

\begin{titlepage}

\thispagestyle{empty}

\begin{flushright}
\begin{tabular}{l}
LAPTH-1199/07
\end{tabular}
\end{flushright}

\vskip1cm

\centerline{\large \bf Anomalous dimensions of leading twist conformal operators}

\vspace{1cm}

\centerline{\sc A.V. Belitsky$^a$, J. Henn$^b$, C. Jarczak$^{c,d}$,
D. M\"uller$^e$, E. Sokatchev$^b$}

\vspace{10mm}

\centerline{\it $^a$Department of Physics and Astronomy, Arizona State University}
\centerline{\it Tempe, AZ 85287-1504, USA}

\vspace{3mm}

\centerline{\it $^b$Laboratoire d'Annecy-le-Vieux de Physique Th\'{e}orique  LAPTH}
\centerline{\it B.P.~110,  F-74941 Annecy-le-Vieux,
France\footnote{UMR 5108 associ\'{e}e {\`a} l'Universit\'{e} de Savoie}}

\vspace{3mm}

\centerline{\it $^c$Institut Fourier, Laboratoire de
Math\'ematiques} \centerline{\it B.P.~74, F-38402 St Martin d'Heres,
France\footnote{UMR 5582 associ\'{e}e {\`a} l'Universit\'e Joseph
Fourier}}

 \vspace{3mm}

\centerline{\it $^d$Laboratoire de Physique, Groupe de Physique
Th\'eorique, ENS Lyon} \centerline{\it46, All\'ee d'Italie, F-69364
Lyon, France\footnote{UMR 5672 associ\'{e}e {\`a} l'Ecole Normale
Sup\'erieure de Lyon}}

 \vspace{3mm}

\centerline{\it $^e$Institut f\"ur Theoretische Physik II,
Ruhr-Universit\"at Bochum} \centerline{\it D-44780 Bochum, Germany}

\vspace{1cm}

\centerline{\bf Abstract}

\vspace{5mm}

We extend and develop a method for perturbative calculations of anomalous dimensions
and mixing matrices of leading twist conformal primary operators in conformal field
theories. Such operators lie on the unitarity bound and hence are conserved (irreducible)
in the free theory. The technique relies on the known pattern of breaking of the
irreducibility conditions in the interacting theory. We relate the divergence of the
conformal operators via the field equations to their descendants involving an extra field
and accompanied by an extra power of the coupling constant. The ratio of the two-point
functions of descendants and of their primaries determines the anomalous dimension,
allowing us to gain an order of perturbation theory. We demonstrate the efficiency of
the formalism on the lowest-order analysis of anomalous dimensions and mixing matrices
which is required for two-loop calculations of the former. We compare these results to
another method based on anomalous conformal Ward identities and constraints from the
conformal algebra. It also permits to gain a perturbative order in computations of mixing
matrices. We show the complete equivalence of both approaches.

\end{titlepage}

\setcounter{footnote} 0

\newpage

\pagestyle{plain} \setcounter{page} 1

%%%%%%%%%%%%%%%%%%%%%%%%%%%%%%%%%%%%%%%%%%%%%%%%%%%%%%%%%%%%%%%%%%%%%
\section{Introduction}
%%%%%%%%%%%%%%%%%%%%%%%%%%%%%%%%%%%%%%%%%%%%%%%%%%%%%%%%%%%%%%%%%%%%%

The framework of the Wilson-Kadanoff operator product expansion \cite{Wil69} for correlation
functions of field operators in quantum field theory, which determine physical observables,
enormously facilitates the analysis of their short- (or light-cone-) distance structure. Since
its discovery it has found a large range of applications stretching from  phase transition
phenomena in condensed matter physics to scattering amplitudes in four-dimensional gauge
theories. The main advantage of the formalism is that it allows one to evaluate a product of
field operators near the short or light-cone distance singularity in terms of composite
operators. In an interacting field theory, the latter mix among each other under renormalization
and acquire nontrivial anomalous dimensions. Their evaluation at higher orders in the coupling
constant is one of the goals of perturbative field theory.

The recent surge of interest in anomalous dimensions of Wilson operators in the maximally
supersymmetric Yang-Mills theory ($\cN=4$ SYM) was inspired by the gauge/string correspondence
which identifies them with the energies of excitations in the dual description in terms of a
string theory on an AdS$_5 \times$S$^5$ background \cite{Mal97}. The strong/weak nature of
this duality makes its straightforward tests difficult, since the perturbative coupling
expansion windows in both theories do not overlap. Due to complications in the quantization
of string theories on warped backgrounds, such direct tests would require an exact evaluation
of anomalous dimensions in gauge theory. In practical terms, what one can realistically do is
to perform multiloop computations, then use results as initial data for the conjectured
integrability\footnote{The literature on the subject is vast, we therefore refer the reader
to comprehensive proceedings of a recent workshop for details,
{\tt http://www-spht.cea.fr/Meetings/Rencitz2007/agenda.php}.} of $\cN=4$ SYM. In this way
one may be able to determine the strong-coupling asymptotics of the anomalous dimensions and
compare it to the string theory predictions.

The goal of the present study is to develop a formalism for
efficient multiloop calculations of anomalous dimensions of a
certain class of Wilson operators in $\cN=4$ SYM, namely operators
of leading twist. An important feature of this model is that it
stays anomaly-free even quantum mechanically and thus preserves all
classical symmetries of the Lagrangian. The central object of our
consideration is a correlation function of two conformal primary
operators of leading twist. The conformal symmetry of the model
completely determines (up to normalization) their functional
dependence on the space-time interval between these points with the
exponent given by the scaling dimension of the composite operators
(given by the sum of their canonical and anomalous dimensions). A
key observation for our formalism is that twist-two conformal
primary operators lie near the conformal unitarity bound and hence
are conserved in the non-interacting theory. However, they acquire a
non-vanishing divergence in the interacting theory. It is obtained
by applying the field equations of motion and thus is proportional
to the  coupling constant. In conformal terms, this divergence
defines a particular conformal descendant of the primary operator.
The idea of the method is to compute the ratio of the two-point
correlation functions of descendants and of their primaries. This
ratio is proportional to the anomalous dimension and involves an
overall factor of two powers of the coupling constant. Thus, in
order to evaluate the anomalous dimension at order $n$ in
perturbation theory, it is sufficient to compute the two correlators
at order $n-1$ and thus gain an order of perturbation theory.

The method we describe here has been first used in \cite{Ans98} to calculate the one-loop
anomalous dimensions of some twist-two operators of low spin in $\cN=4$ SYM and then generalized
to arbitrary spin in \cite{HenJacSok05}. In this paper we present a simpler version of
\cite{HenJacSok05} which does not require supersymmetry. A similar method was applied to obtain
the two- \cite{Ede03} and three-loop \cite{EdeJacSok04} anomalous dimension of the Konishi
operator and of the twist-three operator of the BMN series.

It is a common knowledge that conformal symmetry is broken provided
that the renormalization group function $\beta$ of the coupling
constant is non-vanishing. This is a direct consequence of
dimensional transmutation which generates an intrinsic mass scale in
the theory, modifying the scaling behavior of correlation functions.
In perturbative calculations one has to use a regularization
procedure for ultraviolet divergences to render correlation
functions finite. The only consistent regularization method for
non-Abelian gauge theories is dimensional regularization or its
spin-off, dimensional reduction. However, neither of them preserves
all space-time symmetries of the regularized theory, in particular,
they violate the scaling and special conformal boosts symmetries.
This has profound consequences for the form of the correlation
functions even when the regulator is eliminated. Namely, the
non-zero $\beta-$function in $D = 4-2\varepsilon$ dimensions is
$\beta_{\varepsilon}(g)= - 2 \varepsilon g + \beta (g)$ and it
induces an anomaly in the trace of the energy-momentum tensor. The
renormalization of the product of the renormalized energy momentum
tensor and conformal operators generates anomalous dimensions of the
latter and leads to their mixing under conformal boosts, as we will
demonstrate in this study. Even when the four-dimensional
$\beta-$function is zero to all orders of perturbation theory, the
conformal symmetry is violated for $\varepsilon \neq 0$. Subtracting
divergences and sending $\varepsilon\to 0$ afterwards, one generates
symmetry breaking contributions to the dilatation operator coming
from terms $\sim \beta_{\varepsilon}(g)/ \varepsilon$. This source
of the symmetry breaking is a peculiar feature of dimensional
regularization rather than an intrinsic property of the dilatation
operator. In other words, in gauge theories with vanishing $\beta$
function the conformal symmetry breaking terms can be removed by
performing a scheme transformation of the dilatation operator and by
going over to the so-called conformal scheme. This transformation
does not affect the eigenvalues of the dilatation operator but it
does change the form of the corresponding eigenstates, and is
required for evaluation of correlation functions to preserve their
diagonality.

Our subsequent presentation is organized as follows. In Sect.\ \ref{Sec-GenFor},
we give a general introduction to the method of computing anomalous dimensions
of leading twist operators by differentiation, applicable to any conformal any
field theory. In Sect.\ \ref{Sect-EvaAnoDim}, we begin with a one-loop
calculation of anomalous dimensions of conformal operators in
six-dimensional scalar field theory with
cubic interaction. Then we turn to two-loop order and demonstrate
that conformal symmetry is broken and bare conformal operators start
to mix in the minimal subtraction scheme. We introduce the so-called
conformal scheme which preserves the autonomous renormalization
group equation for conformal operators. We demonstrate the mixing
for the scalar theory making use of an appropriate two-point
correlation function of a conformal operator and its descendant. We
explain how this method allows us to gain an order of perturbation
theory. We then  turn to $\cN=4$ SYM and compute the mixing matrix.
Section \ref{Sect-OriMix} is dedicated to an alternative computation
of the mixing matrix within a different formalism to evaluate the
same quantity making use of conformal Ward identities and commutator
constraints stemming from conformal algebra. Finally, we conclude.

%%%%%%%%%%%%%%%%%%%%%%%%%%%%%%%%%%%%%%%%%%%%%%%%%%%%%%%%%%%%%%%%%%%%%
\section{Anomalous dimensions by differentiation}
\label{Sec-GenFor}
%%%%%%%%%%%%%%%%%%%%%%%%%%%%%%%%%%%%%%%%%%%%%%%%%%%%%%%%%%%%%%%%%%%%%

We are interested in renormalization properties of operators of
leading twist (i.e., dimension minus spin) in $D=2 h$ dimensions and
will restrict ourselves to those that are build from scalar fields.
We are going to outline our method using a simple six-dimensional
model ($\phi^3$ theory),
\begin{equation}
\label{2-1}
\mathcal{L}
=
\sum_{i=1}^{N_{f}}
\left( \partial_{\mu} \phi_{i} \right) \left( \partial^{\mu} \bar{\phi}^{i} \right)
+
\ft{1}{2} \left( \partial_{\mu} \chi \right) \left( \partial^{\mu} \chi \right)
+
g \sum_{i=1}^{N_{f}} \phi_{i} \bar{\phi}^{i} \chi
\, ,
\end{equation}
which is conformally invariant at the classical level. Moreover, its $\beta$ function \cite{MikRad84}
\begin{equation}
\label{2-2}
\beta(g)
= - \beta_{0} \frac{g^3}{(2\pi)^3}
+
O(g^5)
\qquad\,
\beta_{0} = \ft{1}{6} - \ft{1}{24} N_{f}
\end{equation}
can be made vanish up to order $g^5$ in coupling constant by
choosing $N_{f}=4$. This allows us to use conformal symmetry
arguments up to this order of perturbation theory. For the specific
operators that we are interested in, the treatment of $\cN=4$ SYM
requires only minor changes. The $\cN=4$ SYM Lagrangian in Minkowski
space has the form\footnote{We use conventions of Ref.\
\cite{BelDerKorMan03}.}
\begin{eqnarray}
\label{Def-Act-N4}
{\cal L}_{{\cal N} = 4} = {\rm tr} \, \bigg\{
\!\!\!&-&\!\!\! \ft12 F_{\mu\nu} F^{\mu\nu}
+
\ft12 \left( {\cal D}_\mu \phi^{AB} \right) \left( {\cal D}^\mu \phi_{AB} \right)
+
\ft{1}8 g_{\scriptscriptstyle\rm YM}^2 [\phi^{AB}, \phi^{CD}] [\phi_{AB}, \phi_{CD}]
\nonumber\\
&+&\!\!\! 2 i \bar\lambda_{\dot\alpha A} \sigma^{\dot\alpha
\beta}_\mu {\cal D}^\mu \lambda^A_\beta
-
\sqrt{2} g_{\scriptscriptstyle\rm YM} \lambda^{\alpha A} [\phi_{AB}, \lambda_\alpha^B]
+
\sqrt{2} g_{\scriptscriptstyle\rm YM} \bar\lambda_{\dot\alpha A} [\phi^{AB}, \bar\lambda^{\dot\alpha}_B]
\bigg\} \, .
\end{eqnarray}
In the following we perform perturbative expansions with respect to the coupling $g=\sqrt{N_c}
g_{\scriptscriptstyle\rm YM}$.

%%%%%%%%%%%%%%%%%%%%%%%%%%%%%%%%%%%%%%%%%%%%%%%%%%%%%%%%%%%%%%%%%%%%%
\subsection{Preliminaries}
%%%%%%%%%%%%%%%%%%%%%%%%%%%%%%%%%%%%%%%%%%%%%%%%%%%%%%%%%%%%%%%%%%%%%

The composite operators under consideration are built from two field
operators and are generically written as
\begin{equation}\label{2-3}
O^{\mu_{1} \cdots \mu_{j}} = \sum_{k=0}^{j} a_{jk}
\partial^{\{\mu_{1}} \cdots
\partial^{\mu_{k}} \, \varphi \,
\partial^{\mu_{k+1}} \cdots \partial^{\mu_{j}\}}\, \bar{\varphi} \,,
\end{equation}
where $\{ \cdots \}$ stands for traceless symmetrization, and
$\varphi$ stands for an elementary scalar field of canonical
dimension $h-1$ in $D=2 h$ space-time dimensions. In ${\cal N}=4$
SYM they have leading twist two, in the $\phi^3$ theory they have
twist four. From Eq.~(\ref{2-3}) we can see that for a given spin
$j$ there are in general $j+1$ coefficients $a_{jk}$ to determine
(one of which is just an overall normalization).

There exists a continuous series of unitary irreducible representations of the conformal group
characterized by the conformal dimension $d$ and the Lorentz spin $j>0$ (symmetric traceless
tensor). Unitarity requires that
\begin{equation}
\label{2a-1}
d \geq 2(h-1) + j
\, .
\end{equation}
When the bound is saturated, these representations become reducible and one can impose irreducibility
conditions. For the operators $O^{\mu_{1} \cdots \mu_{j}}$ this happens in the free field theory,
where they possess the canonical dimension
\begin{equation}
\label{2a-2}
d_{0,j}=2(h-1) + j
\,.
\end{equation}
Then the irreducibility condition implies that the $O^{\mu_{1} \cdots \mu_{j}}$ are conserved tensors:
\begin{equation}\label{2-11}
\pa_{\mu} O^{\mu \mu_{2} \cdots \mu_{j}} = 0\,.
\end{equation}
Equations (\ref{2-11}) allows one to fix the coefficients $a_{jk}$ in (\ref{2-3}) up to an overall
normalization.

Before writing down the solution, however, let us introduce an efficient tool for traceless
symmetrization of the vector indices \cite{DobPetPetTod76}. It consists in projecting all the
vector indices of a given tensor with a null vector\footnote{In Minkowski space this is a
light-like vector, in Euclidean space it is a complex isotropic vector.}  $z^{\mu}$, $z^2=0$,
thus automatically symmetrizing Lorentz indices and suppressing all the traces. Thus we define
the projected operator
\begin{equation}\label{2-4}
\hat{O}_j
\equiv
O^{\mu_{1} \cdots \mu_{j}} z_{\mu_{1}} \cdots
z_{\mu_{j}}
=
\sum_{k=0}^{j} a_{jk} \hat{\partial}^{k} \,\varphi \,
\hat{\partial}^{j-k}\, \bar{\varphi} \,,
\end{equation}
where we introduced the notation $\hat{\partial} = z^{\mu} \partial_{\mu}$. To free some indices
from contractions with $z^\mu-$vectors (e.g., in order to take a divergence), one has to differentiate
with respect to the auxiliary vector $z^{\mu}$, in the presence of the constraint $z^2=0$. The lowest
order differential operator which does this is second-order and reads\footnote{Owing to $\lbrack
\Delta_{\mu}, z^2 \rbrack = 2 z^2 \frac{\partial}{\partial z^{\mu}} = 0$, this is compatible with
the constraint $z^2 =0$.} \cite{BarTod77}
\begin{equation}\label{2-5}
\Delta^{\mu} = (h-1 + z \cdot \partial_{z})
\partial_{z}^{\mu} - \frac{1}{2} z^{\mu} \partial_{z} \cdot \partial_{z}
\,.
\end{equation}
So, for instance, we may recover the operator in (\ref{2-3}) from the projected one in (\ref{2-4}) by
\begin{equation}\label{2-6}
O^{\mu_{1} \cdots \mu_{j}} = N \, \Delta^{\mu_{1}} \cdots \Delta^{\mu_{j}} \hat{O}_j\,,
\end{equation}
where $N$ is an inessential normalization constant. The symmetrization and tracelessness of
$O^{\mu_{1} \cdots \mu_{j}}$ in (\ref{2-6}) are automatic due to the properties
\begin{equation}
\label{2-7}
\lbrack
\Delta^{\mu},\Delta^{\nu} \rbrack = 0 \,,\qquad \Delta^{\mu}
\Delta_{\mu} = 0
\end{equation} of the differential operator $\Delta^{\mu}$. Let us now rewrite (\ref{2-4}) in a
bi-local form,
\begin{equation}
\label{2-8}
\hat{O}_j = P_{j}\left(\hat{\partial}_{a},\hat{\partial}_{b}\right)
\varphi(x_a) \bar{\varphi}(x_b)
\, ,
\end{equation}
where
\begin{equation}\label{2-9}
P_{j}
\left(x,y\right) = \sum_{k=0}^{j}a_{jk} x^{k}y^{j-k}
\equiv
(x+y)^{j} p_{j}\left( \frac{x-y}{x+y} \right)
\end{equation}
is a homogeneous polynomial of degree $j$.

Using the projection variables $z^\mu$ and the differential operator $\Delta^{\mu}$ introduced
in (\ref{2-5}), we can rewrite (\ref{2-11}) in the following way:
\begin{equation}\label{2-12}
\pa_{\mu} \Delta^{\mu} \hat{O}_j = 0\,.
\end{equation}
The advantage of (\ref{2-12}) over (\ref{2-11}) now becomes evident: the condition (\ref{2-12})
turns into a simple differential equation \cite{TodMinPet78} for the polynomial $p_{j}(x)$
defined in
(\ref{2-9}),
\begin{equation}
\label{2-13}
\left(
(1-x^2)\frac{d^2}{dx^2}-(2\nu+1)x\frac{d}{dx}+j(j+2\nu)
\right)
p_{j}(x) =0\,,
\end{equation}
where\footnote{For fields with a light-cone spin $s$ the index is
$\nu=h-3/2+s$.} $\nu=h-3/2$. Its regular solution is the Gegenbauer
polynomial $C^{\nu}_j(x)$ and hence the operators (\ref{2-8}) read
\cite{Mak81,Ohr82}
\begin{eqnarray}
\label{2-14}
\hat{O}_j = (\hat{\partial}_{a} + \hat{\partial}_{b})^{j} \;
C_{j}^{\nu}
\left(\frac{\hat{\partial}_{a}-\hat{\partial}_{b}}{\hat{\partial}_{a}+\hat{\partial}_{b}}\right)
\varphi(x_a) \bar{\varphi}(x_b)
\, , \quad \nu = h - 3/2\,.
\end{eqnarray}

%%%%%%%%%%%%%%%%%%%%%%%%%%%%%%%%%%%%%%%%%%%%%%%%%%%%%%%%%%%%%%%%%%%%%
\subsection{Anselmi's trick}
%%%%%%%%%%%%%%%%%%%%%%%%%%%%%%%%%%%%%%%%%%%%%%%%%%%%%%%%%%%%%%%%%%%%%

Here we illustrate the main idea of our method \cite{Ans98}. We exploit conformal properties of
quantum field theories supposing that the renormalization has already been done. When the
regulator, e.g., $\varepsilon$ in dimensional regularization, is sent to zero, one expects
conformal properties to emerge. As it is well known, renormalization introduces scheme dependence.
We define our renormalization scheme (within dimensional regularization) such that conformal
primary operators have conformal correlation functions \cite{Mul97}.

When interactions are turned on ($g\neq0$), the scaling dimension of
the tensors $\hat{O}_j$ becomes in general coupling-dependent. In
the following we assume that the quantum numbers of the composite
operators are chosen in such a manner that they form a set closed
under renormalization. After the diagonalization of the mixing
matrix, the operators have well-defined conformal properties. In
particular, their scaling dimension can be simply written as a sum
of the canonical $d_{0,j}$ and anomalous $\gamma_{j}(g^2)$
dimension,
\begin{equation}
\label{3-2a}
d_j=d_{0,j}+\gamma_{j}(g^2)\,.
\end{equation}
Then the unitarity bound (\ref{2a-1}) is no longer saturated and hence the conservation equation
(\ref{2-12}) turns into the non-conservation equation
\begin{equation}
\label{2-16} \partial^{\mu} \Delta_{\mu} \hat{O}_j = g
\hat{K}_{j-1} \,.
\end{equation}
The right-hand side defines a (classical) conformal descendant (a state in the infinite-dimensional
space of the conformal UIR). In our case, where the primary (lowest weight) state is a bilinear
operator, the descendant $\hat{K}_{j-1}$ is trilinear in the fields and can be calculated using the
classical field equations following from (\ref{2-1}).

In a conformal field theory, the form of the two-point function of conformal primary operators
$\hat{O}_j$ of conformal dimension $d_j$ and spin $j$ is fixed to be\footnote{This equation
can be found by requiring covariance under the action of the conformal group, see, e.g.,
\cite{TodMinPet78}.}
\begin{equation}\label{3-1}
\langle \hat{O}_j(x_1)\, \hat{O}_k(x_2) \rangle = \delta_{jk}
C_j(g)\ \hat{I}^{j}\
(x_{12}^2)^{-d_j}
\end{equation}
where
\begin{equation}\label{3-2}
\hat{I} \equiv I_{\mu\nu}z_{1}^{\mu}z_{2}^{\nu}
\, , \qquad
I^{\mu\nu} = \eta^{\mu\nu} - 2 \frac{x_{12}^{\mu}x_{12}^{\nu}}{x_{12}^2}
\, , \qquad
x^{\mu}_{12}=x^{\mu}_{1}-x^{\mu}_{2}
\, ,
\end{equation}
and $C_j(g)$ are normalization constants. In (\ref{3-1}) we have used two independent projection
variables, $z_{1}^{\mu}$ and $z_{2}^{\mu}$, for the operators at points $x^\mu_1$ and $x^\mu_2$,
respectively.

The idea now is to take the divergence at both points of (\ref{3-1}) and to replace the result by
the descendant (\ref{2-16}),\footnote{Equation (\ref{3-5}) is to be understood for non-coincident
points, $x_{1}\neq x_{2}$, otherwise one would need to consider possible contact terms. These
terms are irrelevant for our discussion, so we can simply ignore them.}
\begin{equation}
\label{3-5} \pa^{\mu}_{1}\Delta_{1\mu} \pa^{\nu}_{2}\Delta_{2\nu}
\langle \hat{O}_j(x_1) \,\hat{O}_j(x_2) \rangle = g^2 \langle
\hat{K}_{j-1}(x_1)\, \hat{K}_{j-1}(x_2)\rangle \,.
\end{equation}
Using the expressions (\ref{3-1}), (\ref{3-2}), it is straightforward to carry out
the differentiation in (\ref{3-5}). When it is done, we equate the auxiliary vectors
$z_{1}^{\mu}=z_{2}^{\mu}=z^{\mu}$, for simplicity. Then we evaluate the ratio to be
\begin{eqnarray} \label{3-6}
g^2\,\hat{x}^2
\frac{
\langle \hat{K}_{j-1}(x_1)\, \hat{K}_{j-1}(x_2) \rangle
}{
\langle \hat{O}_j(x_1)\, \hat{O}_j(x_2)
\rangle} &=& -\gamma_{j}(g^2) \,j (j+h-2) \Big[ (j+h-1)(j+2h-3)
\\[-3mm]
&&\hspace{3.5cm} + \gamma_{j}(g^2) \,(j^2+hj-2j+h-1)\Big] \,.
\nonumber
\end{eqnarray}

In a conformal field theory, the ratio (\ref{3-6}) is considered as an exact (non-perturbative)
expression. In practice we can calculate its left-hand side only perturbatively. It is important
to realize the appearance of the factor $g^2$ in the left-hand side of (\ref{3-6}). If we want
to determine
\begin{equation}\label{pertexp}
\gamma_{j}(g^2)
=
\frac{g^2}{(2\pi)^{h}}\ \gamma_{j}^{(0)} + \frac{g^4}{(2\pi)^{2h}}\ \gamma_{j}^{(1)} + \ldots\ ,
\end{equation}
up to, say, order $(g^2)^n$, we only need to evaluate the correlators on the left-hand side of
(\ref{3-6}) up to order $(g^2)^{n-1}$. In other words, this method allows us to gain one perturbative
order in the calculation of $\gamma_{j}(g^2)$. This simple observation is the main point in our
approach.

%%%%%%%%%%%%%%%%%%%%%%%%%%%%%%%%%%%%%%%%%%%%%%%%%%%%%%%%%%%%%%%%%%%%%
\section{Evaluation of anomalous dimensions}
\label{Sect-EvaAnoDim}
%%%%%%%%%%%%%%%%%%%%%%%%%%%%%%%%%%%%%%%%%%%%%%%%%%%%%%%%%%%%%%%%%%%%%

In this section, we use the general formula \re{3-6} to perform actual loop calculations. We start
with a simple scalar field-theory example and then turn to maximally supersymmetric gauge theory,
which is the focus of our study.

%%%%%%%%%%%%%%%%%%%%%%%%%%%%%%%%%%%%%%%%%%%%%%%%%%%%%%%%%%%%%%%%%%%%%
\subsection{Anomalous dimensions in  $\phi^3$ theory in six dimensions}
%%%%%%%%%%%%%%%%%%%%%%%%%%%%%%%%%%%%%%%%%%%%%%%%%%%%%%%%%%%%%%%%%%%%%

%%%%%%%%%%%%%%%%%%%%%%%%%%%%%%%%%%%%%%%%%%%%%%%%%%%%%%%%%%%%%%%%%%%%%
%            Figure
%%%%%%%%%%%%%%%%%%%%%%%%%%%%%%%%%%%%%%%%%%%%%%%%%%%%%%%%%%%%%%%%%%%%%
\begin{figure}[t]
\begin{center}
\mbox{
\begin{picture}(0,155)(205,0)
\put(0,0){\insertfig{14.5}{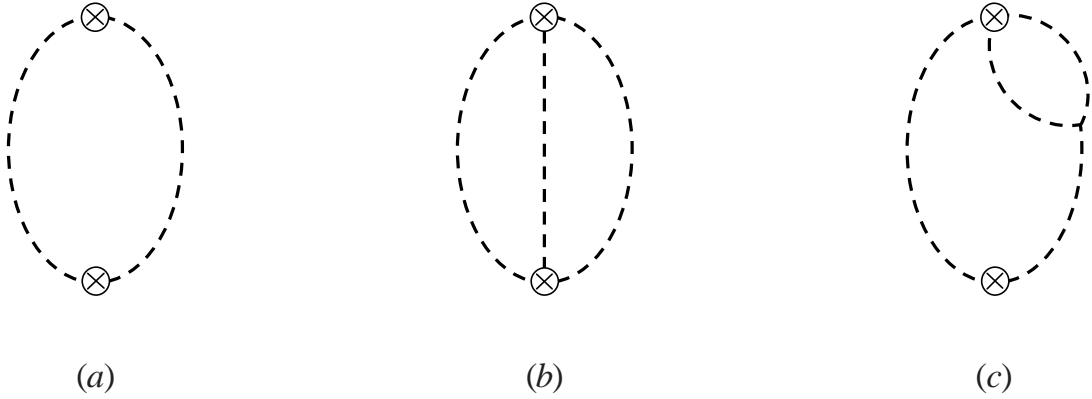}}
\end{picture}
}
\end{center}
\caption{\label{FigCorFunSca} Feynman diagrams  for the evaluation of anomalous dimensions ($a$
and $b$) and mixing matrix ($c$) in $\phi^3$ theory. The symbol $\otimes$ stands for the conformal
operator.}
\end{figure}%
%%%%%%%%%%%%%%%%%%%%%%%%%%%%%%%%%%%%%%%%%%%%%%%%%%%%%%%%%%%%%%%%%%%%%

Let us consider conformal operators of leading twist,
\begin{eqnarray}
\label{Def-Ope-phi3}
\hat{O}_j = (\hat{\partial}_{a} + \hat{\partial}_{b})^{j} \;
C_{j}^{3/2}
\left(\frac{\hat{\partial}_{a}-\hat{\partial}_{b}}{\hat{\partial}_{a}+\hat{\partial}_{b}}\right)
\phi_1(x_a) \bar{\phi}^2(x_b) \,,
\end{eqnarray}
which are `flavor' non-singlets (to avoid additional mixing) and are thus closed under
renormalization\footnote{The case $j=1$ corresponds to the one of the conserved $U(N_{f})$
currents, $O^{\mu} = \pa^{\mu} \phi_{1} \bar{\phi}^{2} - \phi_{1}\pa^{\mu}  \bar{\phi}^{2}$, for
which $\gamma_{j=1}(g^2)=0$.}. The descendants (\ref{2-16})  are straightforwardly evaluated
and are given by
\begin{equation}\label{2-17}
\hat{K}_{j-1} = (\hat{\pa}_{a} + \hat{\pa}_{b} + \hat{\pa}_{c})^{j-1} k_{j-1}\left(
\frac{\hat{\pa}_{a} + \hat{\pa}_{b} - \hat{\pa}_{c}}{\hat{\pa}_{a} + \hat{\pa}_{b} + \hat{\pa}_{c}},
\frac{- \hat{\pa}_{a} + \hat{\pa}_{b} - \hat{\pa}_{c}}{\hat{\pa}_{a} + \hat{\pa}_{b} + \hat{\pa}_{c}}
\right) \chi(x_a) \phi_{1}(x_b) \bar{\phi}^{2}(x_c) \,, \end{equation}
where
\begin{equation}\label{2-18}
k_{j-1}(x,y) =  3 \left[ 2 C^{5/2}_{j-1}(x) - 5 (1-x)
C^{7/2}_{j-2}(x) - 2 C^{5/2}_{j-1}(y) - 5 (1+y) C^{7/2}_{j-2}(y)
\right]
\end{equation}
is now a polynomial of two variables rather than of one.

We now utilize the ratio (\ref{3-6}) of two-point functions of primaries and descendants to
evaluate the anomalous dimension.  Since we gained a factor of $g^2$, it is clear that at
lowest order we have
\begin{equation}\label{3-6a}
(2\pi)^{h} \hat{x}^2
\frac{
\langle \hat{K}_{j-1}(x_1)\, \hat{K}_{j-1}(x_2) \rangle|_{g^0}
}{
\langle \hat{O}_j(x_1)\, \hat{O}_j(x_2)
\rangle|_{g^0}}
= -\gamma_{j}^{(0)}\, j (j+h-2) (j+h-1)
(j+2h-3) \,,
\end{equation}
where the left-hand side can be evaluated by a tree-level (order $g^{0}$) calculation. Depending
on whether we want to use this formula in the $\phi^3$ model or in $\cN=4$ SYM, we can specialize
it to either six ($h=3$) or four ($h=2$) dimensions to determine the corresponding leading order
anomalous dimensions.

A comment on the evaluation of the two point functions that arise in (\ref{3-6a}) is in order here.
Although it is a tree level calculation, the spin dependence presents a technical difficulty.
Namely, the composite operators $\hat{O}_j,\ \hat{K}_{j-1}$ contain polynomials in the projected
derivatives (see, e.g., (\ref{2-18})), and expanding them one gets a result for their two-point
functions in terms of multiple sums. Therefore, it is more convenient to use the Schwinger
representation for the scalar Euclidean propagators,
\begin{equation}\label{3-7}
\langle \phi_{i}(x_1) \bar{\phi}^{j}(x_2)\rangle =  \frac{\Gamma(h-1)}
{4\pi^{h}}\frac{\delta_i^j}{(x_{12}^2 )^{h-1}} = \delta_i^j\int_{0}^{\infty}
\frac{d\alpha}{4\pi^{h}}\, \alpha^{h-2} e^{-\alpha x_{12}^2}\,,
\end{equation}
so that all projected derivatives acting on propagators are
essentially replaced by the $\alpha-$pa\-ra\-me\-ters. As a
consequence, instead of infinite sums one obtains integrals over the
polynomials that appear in the definition of the conformal
operators, see, e.g., (\ref{2-14}), (\ref{2-18}), which are then
evaluated using standard properties of the Gegenbauer polynomials,
in particular their orthogonality relation
\begin{equation}\label{A2-3}
\int_{0}^{1}dx \left[x (1-x) \right]^{\nu-1/2} C^{\nu}_{j}(x)
C^{\nu}_{k}(x) = \delta_{jk} \frac{2^{1-4\nu}  \pi
\Gamma(j+2 \nu)}{\Gamma^{2}(\nu) \Gamma(j+1) (j+\nu)} \,.
\end{equation}

Let us now use (\ref{3-6a}) (with $h=3$) in order to calculate the one-loop anomalous dimensions
of the twist-four operators (\ref{2-14}). We need to evaluate the two correlators at order $g^0$,
see Fig.~\ref{FigCorFunSca} (a). The first one fixes the normalization:
\begin{eqnarray}\label{3-9}
\langle \hat{O}_j(x_1) \hat{O}_k(x_2) \rangle
&=&
\frac{(- 2\hat{x}_{12})^{j+k}}{(4\pi^3)^2}  \int_{0}^{\infty}
d\alpha \, \alpha \int_{0}^{\infty} d\beta \, \beta \, (\alpha +
\beta)^{j+k} \nonumber
\\
&& \times \, C^{3/2}_{j}\left(\frac{\alpha-\beta}{\alpha +
\beta}\right) \, C^{3/2}_{k}\left(\frac{\alpha-\beta}{\alpha +
\beta}\right) \, {\rm exp}{\left[-(\alpha+\beta)x^{2}_{12}
\right]}  \,.
\end{eqnarray}
The integrals with respect to Schwinger parameters are
straightforwardly evaluated using (\ref{3-7}) and yield the
following result:
\begin{equation}\label{3-9a}
\langle \hat{O}_j(x_1) \hat{O}_k(x_2) \rangle
= \delta_{jk} (j+1)(j+2)(2j+2)! \,  \frac{2^{2j-2}}{(4\pi^3)^2}
\frac{( \hat{x}_{12} )^{2j}}{(x_{12}^2)^{2j+4}} \, .
\end{equation}
It indeed possesses the orthogonal form (\ref{3-1}) expected on the basis
of conformal symmetry. The correlation function for descendants, i.e.,
$\langle \hat{K}_{j-1}(x_1) \hat{K}_{k-1}(x_2)\rangle$, is obtained in a
similar fashion upon evaluation of the Feynman diagram in Fig.~\ref{FigCorFunSca}
(b). Its expression differs from Eq.\ (\ref{3-9a}) only by a factor and thus
the ratio (\ref{3-6a}) gives the well-known leading order approximation for the
anomalous dimensions \cite{Kub79,MikRad84}:
\begin{equation}
\label{3-11}
\gamma_{j}^{(0)}
= \frac{1}{4} \left[ \frac{1}{6}- \frac{1}{(j+1)(j+2)}\right]
\,.
\end{equation}

%%%%%%%%%%%%%%%%%%%%%%%%%%%%%%%%%%%%%%%%%%%%%%%%%%%%%%%%%%%%%%%%%%%%%
\subsection{Scheme ambiguities}
\label{renCFT}
%%%%%%%%%%%%%%%%%%%%%%%%%%%%%%%%%%%%%%%%%%%%%%%%%%%%%%%%%%%%%%%%%%%%%

Up to now we have been able to avoid addressing renormalization
issues because our method of calculating $\gamma_{j}^{(0)}$ involved
only tree-level calculations. However, by going higher up in
perturbation theory, we need to discuss how the composite operators
(\ref{2-14}) are renormalized. As pointed out above, it is
misleading to assume from the start that conformal symmetry is
preserved and an ad hoc renormalization scheme will yield diagonal
correlation functions. The Poincar\'e invariance alone imposes
rather weak constrains on the mixing of operators: an operator with
a given spin $j$ will mix under renormalization with total
derivatives of lower spin operators, such as
\begin{equation}
\label{4-0}
\hat{O}_{jl} = \hat{\pa}^{l-j} \hat{O}_j\qquad{\rm (no}\;{\rm summation).}
\end{equation}
The bare operators $\hat{O}_{jk}$, defined in terms of bare field operators, will be our basis
states to discuss operator mixing. The renormalized operator is obtained as a superposition given
by the renormalization matrix $\mathbb{Z}$,
\begin{equation}\label{4-1}
\hat{\mathbb{O}}_{j}=\sum_{k=0}^{j} {\mathbb{Z}}_{jk} \hat{O}_{kj}\,.
\end{equation}
The mixing pattern implied by Poincar\'e symmetry results in a lower triangular matrix $\mathbb{Z}$,
i.e., $\mathbb{Z}_{jk} = 0$ for $k > j$.  Our goal is to define a conformal scheme, that is, to
choose ${\mathbb{Z}}$ such that the renormalized operators $\hat{\mathbb{O}}_j$ have conformal
two-point functions (cf. (\ref{3-1}))
\begin{equation}\label{4-1a}
\langle \hat{\mathbb{O}}_j(x_1)\, \hat{\mathbb{O}}_k(x_2) \rangle = \delta_{jk}
C_j(g)\ \hat{I}^{j}\
(x_{12}^2)^{-d_j(g)}\,.
\end{equation}
In Eq.\ (\ref{4-1a}) it is understood that the regulator has been removed, $\varepsilon \rightarrow 0$
in dimensional regularization.

We determine the $\mathbb{Z}$ matrix from a calculation in the $\overline{\rm MS}$ scheme and then
perform an additional finite scheme transformation. The rotation matrix is governed by the form
(\ref{4-1a}) of the two-point correlation function. In the $\rm \overline{MS}$ scheme we define the
renormalized operator insertion
\begin{equation}
\label{Def-RenOpeIns}
[\hat{O}_{j}]=\sum_{k=0}^{j} {Z}_{jk} \hat{O}_{kj}
\end{equation}
in terms of a renormalization matrix $Z$. Perturbatively, it is given by the Laurent series:
\begin{eqnarray}
Z_{jk}
=
\delta_{jk}
+
 \sum_{n=1}^{\infty}\frac{g^{2n}}{(2\pi)^{n h}}\sum_{m=1}^{n} \frac{Z^{[m](n)}_{jk}}{\varepsilon^{m}}\,.
\end{eqnarray}
The anomalous dimension matrix is obtained directly from a scale variation applied to Eq.\
(\ref{Def-RenOpeIns}). Since the bare operator does not depend on the renormalization scale
$\mu$ we get the standard relation
\begin{equation}\label{4b-1}
\gamma_{jk}(g^2) =-\lim_{\varepsilon\to 0}\, \mu\frac{d}{d\mu}( \ln Z)_{jk}(g^2)
\end{equation}
for computation of anomalous dimension matrix from the renormalization $Z$-matrix. The perturbative
expansion of the anomalous dimension matrix is analogous to that in Eq.~(\ref{pertexp}). Let us also
recall that in this scheme it is entirely determined by the residue of the $Z$-matrix,
\begin{equation}
\gamma_{jk}(g^2) = g\frac{\partial}{\partial g} Z_{jk}^{[1]}(g^2)
\, ,
\end{equation}
while all higher order poles are fixed  from the renormalizability of the composite operators.
For instance, up to order $g^4$ we have (for a vanishing $\beta$ function):
\begin{equation}\label{4-8}
Z^{[1](1)}_{jk} = \delta_{jk} \frac{1}{2}\gamma^{(0)}_j
\,,\qquad
Z^{[2](2)}_{jk} =  \delta_{jk} \frac{1}{8} \left(\gamma^{(0)}_j\right)^{2}
\,,\qquad
Z^{[1](2)}_{jk} = \frac{1}{4} \gamma^{(1)}_{jk}
\,.
\end{equation}
Beyond leading order approximation, the anomalous dimension matrix in the
scheme is non-diagonal and has a triangular form. Note that the eigenvalues of the anomalous
dimension matrix are given by the diagonal entries and coincide with the scale dimensions of
the conformal operators in the conformal scheme. The scheme transformation to the latter from
$\overline{\rm MS}$ is given by
\begin{equation}
\label{4-6a}
\hat{\mathbb{O}}_{j j}
=
\sum_{k=0}^j B^{-1}_{jk} [\hat{O}_{jk}]\,, \qquad \mathbb{Z}_{jk}=\sum_{m=k}^j B^{-1}_{jm}Z_{mk}
\,.
\end{equation}
The finite renormalization matrix $B$ admits the perturbative series
representation
\begin{equation}
B_{jk} = \delta_{jk} + \sum_{n=1}^{\infty} \frac{g^{2n}}{(2\pi)^{n h}} B^{(n)}_{jk}
\,,
\label{4-7}
\end{equation}
with the expansion coefficients $B^{(n)}_{jk}$ being triangular matrices. In a conformal field
theory where the $\beta-$function vanishes, the  anomalous dimension matrices in the two schemes
are simply related by
\begin{eqnarray}
\label{Dia-AD}
\gamma_j(g^2) \delta_{jk}= \left(B^{-1} \gamma B\right)_{jk}(g^2)\,.
\end{eqnarray}
Hence, the  $B-$matrix diagonalizes the anomalous dimension matrix evaluated in the $\overline{\rm MS}$
scheme. In particular, to the first nontrivial order at which the mixing phenomena occurs we have:
\begin{equation}
\label{Res-B2AD}
\gamma_{jk}^{(1)} = -\left(\gamma_{j}^{(0)} -\gamma_{k}^{(0)}\right) B^{(1)}_{jk}\,.
\end{equation}
Note that the knowledge of $\gamma^{(0)}_j$, and $B^{(1)}_{jk}$ is sufficient to reconstruct the
off-diagonal entries $\gamma_{jk}^{(1)}$  up to order $g^4$.

Let us comment on the orthogonality of conformal operators in non-integer space-time dimensions.
According to Eq.\ (\ref{2-14}), the index of the Gegenbauer polynomials is shifted by $-\varepsilon$.
In practical calculations, we find it convenient to use conformal covariance of the bare operator
in $2 h$ dimensions and therefore we define an ``$\varepsilon-$deformed" basis
\begin{equation}\label{4-0b}
\hat{O}^{(\varepsilon)}_j = (\hat{\partial}_{a} + \hat{\partial}_{b})^{j} \;
C_{j}^{3/2-\varepsilon}
\left(\frac{\hat{\partial}_{a}-\hat{\partial}_{b}}{\hat{\partial}_{a}+\hat{\partial}_{b}}\right)
\phi_{1}(x_a) \bar{\phi}^{2}(x_b)
\,.
\end{equation}
Then instead of (\ref{4-0}) we have
\begin{equation}\label{4-0d}
\hat{\mathbb{O}}_j=\sum_{k=0}^{j} {\mathbb{Z}}^{'}_{jk} \hat{O}^{(\varepsilon)}_{jk}\,.
\end{equation}
The practical advantage of the basis states (\ref{4-0b}) is that the two-point functions at order
$g^{0}$ are orthogonal even in the presence of the regulator $\varepsilon$,
\begin{equation}\label{4-0c}
\langle \hat{O}^{(\varepsilon)}_j(x_1) \hat{O}^{(\varepsilon)}_k(x_2)\rangle|_{g^0} = 0
\, ,
\qquad j\neq k\,.
\end{equation}
It is easy to recover the mixing matrix in the standard basis from the $\varepsilon$-deformed one
by expanding the Gegenbauer polynomials with respect to their index. We define
\begin{equation}\label{A2-1}
\frac{\partial}{\partial \rho} C_{j}^{\nu+\rho}(x)|_{\rho=0}
=
- 2 \sum_{k=0}^{j} d^{\nu}_{jk} C^{\nu}_{k}(x)\,,
\end{equation}
and have for $j>k$ the entries
\begin{equation}\label{A2-2}
d^{\nu}_{jk} = - (1+(-1)^{j-k}) \frac{k+\nu}{(j+k+2\nu)(j-k)} \,.
\end{equation}
Then the desired relation reads
\begin{equation}\label{7a-4}
B^{(1)}_{jk} =B^{'(1)}_{jk} -  d^{(h-1/2)}_{jk} \left( \gamma_j^{(0)} -
\gamma_k^{(0)} \right)\,.
\end{equation}

%%%%%%%%%%%%%%%%%%%%%%%%%%%%%%%%%%%%%%%%%%%%%%%%%%%%%%%%%%%%%%%%%%%%%
\subsection{Operator mixing: perturbative calculations}
\label{opmixpert}
%%%%%%%%%%%%%%%%%%%%%%%%%%%%%%%%%%%%%%%%%%%%%%%%%%%%%%%%%%%%%%%%%%%%%

As we explained in the previous section, the conformal
renormalization scheme is defined by the requirement that the
correctly renormalized conformal primary operator
$\hat{\mathbb{O}}_j$ should lead to the diagonal two-point
correlation functions (\ref{4-1a}). This fixes the renormalization
matrix $\mathbb{Z}$ or, equivalently, $\mathbb{Z}^{'}$. In this
section we show how to determine $B^{'(1)}$ (which, due to
(\ref{7a-4}), is equivalent to finding $B^{(1)}$), by a calculation
at order $g^1$.

For this purpose it is sufficient to employ Eq.~(\ref{4-1a}) for $j>k$,
\begin{equation}\label{5-2a}
\langle \hat{\mathbb{O}}_j(x_1)\, \hat{\mathbb{O}}_k(x_2) \rangle = 0 \,, \qquad  \;j>k\,.
\end{equation}
Let us now expand (\ref{5-2a}) up to order $g^2$ in the coupling constant. With the help of the
expansion
\begin{equation}\label{5-1}
\hat{\mathbb{O}}_j
=
\hat{O}^{(\varepsilon)}_j
+
\frac{g^2}{(2\pi)^{h}} \frac{1}{2 \varepsilon}\gamma^{(0)}_j \hat{O}^{(\varepsilon)}_j
-
\frac{g^2}{(2\pi)^{h}} \sum_{k=0}^{j} B^{'(1)}_{jk} \hat{O}^{(\varepsilon)}_{jk} +O(g^4)
\,,
\end{equation}
we obtain
\begin{equation} \label{5-3}
\langle
\hat{O}_j(x_1)\, \hat{O}_k(x_2) \rangle|_{g^2} - \frac{g^2}{(2\pi)^{h}}
B^{'(1)}_{jk} \langle
\partial^{j-k} \hat{O}_k(x_1)\, \hat{O}_k(x_2)
\rangle|_{g^0} = 0\,, \qquad  \;j>k\,.
\end{equation}
Here we have used the orthogonality relation (\ref{4-0c}) of the operators ${\hat{O}}^{(\varepsilon)}_j$
at tree level, which is why the $1/\varepsilon$ terms from (\ref{5-1}) have also disappeared. The finite
part of (\ref{5-3}) determines in principle $B^{'}_{1}$. However, we can gain an order in $g$ by taking
the divergence $\Delta_{\mu} \partial^{\mu}$ at point $x_1$:
\begin{equation}\label{5-4}
\frac{g^2}{(2\pi)^h} B^{'(1)}_{jk}\,  \langle \partial_{\mu} \Delta^{\mu}
\partial^{j-k} \hat{O}_k(x_1)\, \hat{O}_k(x_2)
\rangle|_{g^0} = g\langle \hat{K}_{j-1}(x_1)\, \hat{O}_k(x_2)
 \rangle|_{g^1} \,, \qquad  \;j>k\,.
\end{equation}
In Eq.\ (\ref{5-4}) the order $O(g)$ correlator has still to be regularized but it turns out to be finite
(up to a contact term) when the regulator is set to zero, see Eq.\ (\ref{5M-1}). Notice that we cannot
gain a further order in $g$ by taking the divergence at the second point in (\ref{5-4}) because then the
equation would be trivially satisfied.

We are going to determine $B^{'(1)}$ from Eq.~(\ref{5-4}). On the right-hand side of this equation we need
to evaluate $\langle \hat{K}_{j-1}(x_1)\, \hat{O}_k(x_2) \rangle|_{g}$ for $j>k$, see Fig.~\ref{FigCorFunSca}
(c). It turns out that the same calculation, but for $j<k$ and $j=k$ provides a useful consistency check.
Firstly, the triangularity of the matrix ${B^{'}}$  requires
\begin{equation}\label{5-9}
\langle \hat{K}_{j-1}(x_1)\, \hat{O}_k(x_2)
 \rangle|_{g}= 0\,,  \qquad  \;j<k\,.
\end{equation}
Secondly, for $j=k$, it follows from (\ref{4-1a}) that
\begin{equation} \label{5-8}
g \,\hat{x}^2 \frac{\langle \hat{K}_{j-1}(x_1)\, {\hat{O}}_j(x_2) \rangle|_{g} }
{\langle \hat{O}_j(x_1)\,
\hat{O}_j(x_2)
\rangle|_{g^0}} = -\frac{g^2}{(2\pi)^h} \gamma^{(0)} j(j+h-2) \,.
\end{equation}
Of course, we could also have taken another divergence at the second
point of the correlator, as in (\ref{3-6a}), to gain a further order
in $g$, but our goal here is to provide a check for the calculation
of the mixing matrix. We have indeed verified that (\ref{5-8})
reproduces the previously determined values (\ref{3-11}),
(\ref{3-14}) of $\gamma^{(0)}$ in the $\phi^3$ theory and in $\cN=4$
SYM, respectively, and that the triangularity condition (\ref{5-9})
is satisfied.

The calculation of $\langle \hat{K}_{j-1}(x_1)\, \hat{O}_k(x_2)
\rangle|_{g}$ involves two Feynman diagrams, see Fig.
\ref{FigCorFunSca} (c), with just one interaction vertex in
$x$-space. The only Feynman integral that arises,
\begin{equation}\label{5M-1}
\int\! d^{6-2\varepsilon} x_{3}\;
\frac{1}{(x_{13}^{2})^{4-2\varepsilon}(x_{23}^{2})^{2-\varepsilon}} =
-\frac{{\pi}^3}{3 (x_{12}^{2})^{3}}  + O(\varepsilon)\,,\qquad x_{1}
\neq x_{2}\,
\end{equation}
is finite in dimensional regularization, up to a contact term. Recall that we always keep $x_{12}\neq 0$
and so we can drop contact terms. Taking the ratio of $\langle \hat{K}_{j-1}(x_1)\, \hat{O}_k(x_2)
\rangle|_{g}$ and $\langle \hat{O}_{j}(x_1)\, \hat{O}_k(x_2) \rangle|_{g^0}$ according to Eq.\ (\ref{5-4}),
we find
\begin{equation}\label{5M-3}
B^{'(1)}_{jk} = \gamma_j^{(0)} d^{3/2}_{jk} \,.
\end{equation}
Switching back to the undeformed basis using (\ref{7a-4}), we have the result :
\begin{equation}\label{5M-4}
B^{(1)}_{jk} = d^{3/2}_{jk} \gamma_k^{(0)}\,,
\end{equation}
which coincides with both the explicit evaluation \cite{MikRad84} and the conformal symmetry predictions
\cite{Mul91}, see Eq.~(\ref{Res-B2AD}).

Historically, there were attempts to determine the form of the renormalized operator insertion by a
shift of the canonical dimensions in the Gegenbauer polynomials \cite{CraDobTod83,MikRad84}. We find
that the renormalized conformal operator insertion can be written to this order as
\begin{eqnarray}
\hat{\mathbb{O}}_j
=
\left(1 + \frac{g^2}{(2\pi)^3} \frac{\gamma_j^{(0)}}{2\varepsilon} \right) (\hat{\partial}_{a}
+
\hat{\partial}_{b})^{j} \; C_{j}^{3/2-\varepsilon+ \gamma_j(g)/2 }
\left(\frac{\hat{\partial}_{a}-\hat{\partial}_{b}}{\hat{\partial}_{a}+\hat{\partial}_{b}}\right)
\phi_1(x_a) \bar{\phi}^2(x_b) + O(g^4)
\, .
\end{eqnarray}
The result we obtain in $\phi^3$ theory, starting with a conformal
invariant operator in $6-2\varepsilon$ dimensions, indeed exhibits
the expectation that the index $\nu= h-3/2$ of the Gegenbauer
polynomials will be shifted by the amount of the anomalous
dimensions, however, it contains explicitly the regularization
parameter contrary to previous proposals. In gauge field theories
such a simple recipe does not work \cite{BroDamFriLep84}, as we will
demonstrate once more in the next section.

%%%%%%%%%%%%%%%%%%%%%%%%%%%%%%%%%%%%%%%%%%%%%%%%%%%%%%%%%%%%%%%%%%%%%
\subsection{Application of the method to $\cN=4$ SYM}
\label{Sec-N4SUSY}
%%%%%%%%%%%%%%%%%%%%%%%%%%%%%%%%%%%%%%%%%%%%%%%%%%%%%%%%%%%%%%%%%%%%%

%%%%%%%%%%%%%%%%%%%%%%%%%%%%%%%%%%%%%%%%%%%%%%%%%%%%%%%%%%%%%%%%%%%%%
%            Figure
%%%%%%%%%%%%%%%%%%%%%%%%%%%%%%%%%%%%%%%%%%%%%%%%%%%%%%%%%%%%%%%%%%%%%
\begin{figure}[t]
\begin{center}
\mbox{
\begin{picture}(0,155)(205,0)
\put(0,0){\insertfig{14.5}{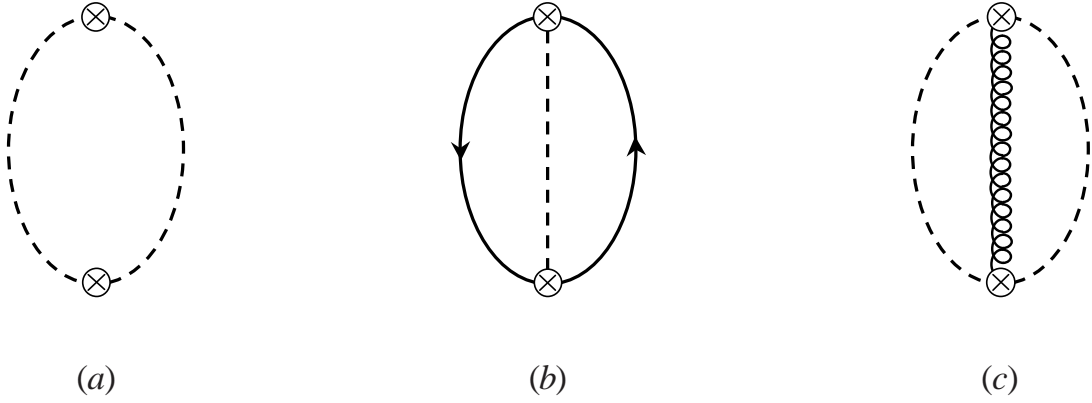}}
\end{picture}
}
\end{center}
\caption{ \label{FigCorFunSYM} Feynman diagrams for the evaluation of anomalous
dimensions in maximally supersymmetric gauge theory.}
\end{figure}%
%%%%%%%%%%%%%%%%%%%%%%%%%%%%%%%%%%%%%%%%%%%%%%%%%%%%%%%%%%%%%%%%%%%%%

In $\cN=4$ SYM one builds twist-two operators by taking bi-linear combinations of the elementary
fields $\phi_{AB},\lambda_{A},\bar{\lambda}^{A}$ and $F^{\mu\nu}$. There are many different
possibilities for constructing such operators \cite{BelDerKorMan03,Bei04}, and in general one faces
an additional mixing problem between scalars, fermions and gluons. This happens for instance for
the superconformal primary operators constructed in \cite{BiaHesRic05,HenJacSok05}. There the
mixing problem at tree level was resolved by exploiting a super-conservation condition required by
superconformal symmetry. Here we avoid this additional mixing problem by considering a different
member of this superconformal multiplet, which is a conformal primary operators in the $\mathbf{20'}$
of $SU(4)$ \cite{BelDerKorMan03,BiaEdeRosSta02} (in which we choose the highest-weight projection of the $SU(4)$
indices of the scalar fields $\phi^{AB}$):
\begin{equation}\label{3-12}
\hat{O}_j = (\hat{\partial}_{a} + \hat{\partial}_{b})^{j} \;
C_{j}^{1/2}
\left(\frac{\hat{\partial}_{a}-\hat{\partial}_{b}}{\hat{\partial}_{a}+\hat{\partial}_{b}}\right)
\rm{tr}  \phi^{12}(a) \phi^{12}(b) \,.
\end{equation}
Note that $j$ has to be even in (\ref{3-12}), and that $j=0$ is a
special case, for which $\hat{O}_{j=0}$ is itself the superconformal
primary operator of the protected energy-momentum supermultiplet.
For $j\neq0$, the $\hat{O}_j$ can be obtained by acting with four
supercharges \cite{BelDerKorMan03} on the superconformal primary
operators of \cite{BiaHesRic05,HenJacSok05}. The fact that these
operators belong to the same supermultiplet implies that their
anomalous dimensions are given by the same universal formula.

The expression of the descendants $\hat{K}_j$ is given in the Appendix, see Eq.\ (\ref{A1-6}).
Schematically, they read
\begin{eqnarray}
\hat{K}_{j-1} &=& A_{j}(\hat{\pa}_{a},\hat{\pa}_{b},\hat{\pa}_{c})
\; \mathrm{tr}\; \left( \{ \lambda^{\alpha}_{3}(x_a),
\lambda_{\alpha 4}(x_c)\} \phi^{12}(x_b) +\{
\bar{\lambda}_{\dot{\alpha}}^{1}(x_a), \bar{\lambda}^{\dot{\alpha}
2}(x_c)\}
\phi^{12}(x_b) \right)  \nonumber \\
&+& B_{j}(\hat{\pa}_{a},\hat{\pa}_{b},\hat{\pa}_{c}) \;
\mathrm{tr}\; {}[
\partial^{\mu} \phi^{12}(x_a), \phi^{12}(x_b) ] F^{\mu\nu}(x_c) z_{\nu}
+ O(g)\,,
\end{eqnarray}
Let us first calculate $\gamma_{j}^{(0)}$ from (\ref{3-6a}). The
calculations presented here were done in the light cone gauge,
i.e.,\ $z\cdot A =0$. However, because the correlators we calculate
are gauge invariant, our results do not depend on the gauge choice.
We have explicitly checked that a calculation in the standard
covariant gauge leads to an identical result. For the $\langle
\hat{K}_{j-1}(x_{1}) \,\hat{K}_{j-1}(x_{2})\rangle$ correlator,
there are two contributing Feynman diagrams, one involving fermions
in Fig.\ \ref{FigCorFunSYM} (b) and the other one gluons in see
Fig.\ \ref{FigCorFunSYM} (c). Their respective contributions to
$\gamma_{j}^{(0)}$ are
\begin{equation}\label{3-13}
\gamma_{j}^{(0)}|_{\rm Fig.\; 2 (a)} = 2 \,,\quad \gamma_{j}^{(0)}|_{\rm Fig.\;
2 (b)} = 2 \left( S_{j}-1 \right)\,,
\end{equation}
where $S_{j} = \sum_{k=1}^{j}1/k$ is a harmonic sum. Adding the two, we obtain the well-known
result
\begin{equation}\label{3-14}
\gamma_{j}^{(0)}= 2 S_{j}\,.
\end{equation}

\begin{figure}[t]
\begin{center}
\mbox{
\begin{picture}(0,155)(205,0)
\put(0,0){\insertfig{14.5}{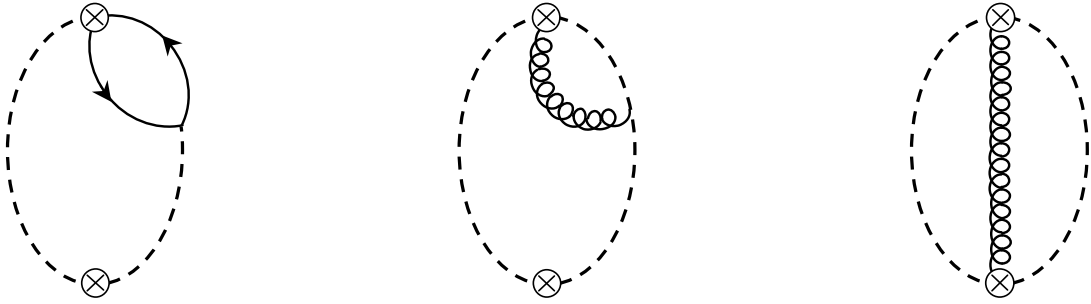}}
\end{picture}
}
\end{center}
\caption{ \label{FigCorFunSYM2} Feynman diagrams needed for the evaluation of the mixing
matrix in $\mathcal{N} = 4$ SYM.}
\end{figure}

Now turning to the mixing matrix, in the left-hand side of (\ref{5-4}), the prefactor
of $B^{(1)}$ can be easily evaluated by a tree-level calculation,
\begin{eqnarray}\label{5-5}
\pa_{1}^{\mu}\Delta_{1 \mu} \partial^{j-k} \langle \hat{O}_{k}(x_1)\,
\hat{O}_{k}(x_2) \rangle_{g^0} &\equiv&  - \frac{1}{d^{1/2}_{jk}}
M_{jk}(x_{12})\,,
\end{eqnarray}
where
\begin{equation}
M_{jk}(x_{12}) =2 (j+k+1)! \frac{(N_{c}^{2}-1)}{(4\pi^2)^2}
\frac{1}{x^6} \left(-2 \frac{\hat{x}^2}{x^2}\right)^{j+k-1}\,,
\end{equation}
and $d^{1/2}_{jk}$ is defined in (\ref{A2-2}). In order to evaluate the right-hand
side of (\ref{5-4}) we need to compute the correlator $\langle \hat{K}_{j-1}
\hat{O}_{j} \rangle|_{g}$ (see Fig.~\ref{FigCorFunSYM2}). We obtain
\begin{equation}\label{7-3}
\langle \hat{K}_{j-1}(x_1)\,\hat{O}_k(x_2) \rangle = M_{jk}(x) \left(
I_{jk} + J_{jk} \right)\,,
\end{equation}
where the integrals $I_{jk},J_{jk}$ come from Fig.\ \ref{FigCorFunSYM2} (a) and the sum of
Figs.\ \ref{FigCorFunSYM2} (b,c), respectively. They are defined by (for $j,k$ even)
\begin{equation}
\label{7-5}
I_{jk} = 2 \int_{-1}^{1} dt\,(1+t)
a_{j}(t) C^{1/2}_{k}(t) =  -2 \qquad (j>k)\,,
\end{equation}
and
\begin{equation}\label{7-9}
J_{jk} =\int_{-1}^{1} ds\, \int_{-1}^{s} dt\, (1+t)\, b_{j}(s,t) \,
C_{k}^{1/2}(s) +\int_{-1}^{1} dt\,\frac{1+t}{1-t}\, \int_{t}^{1}ds\,
\,(1-s) {b}_{j}\left(s,t \right) C^{1/2}_{k}(t)\,.
\end{equation}
The expressions for $a_{j}(t)$ and $b_{j}(s,t)$ can be found in (\ref{A1-8}). With the method
described in the appendix of \cite{BelMue98c} one can evaluate $J_{jk}$, the result being
\begin{eqnarray}\label{7-11}
J_{jk} &=& 2 \left( 2 S_{j-k} +  S_{\frac{j+k}{2}} - S_{\frac{j-k}{2}} - 2 S_{j} +1 \right)
\qquad (j>k)\,.
\end{eqnarray}
Combining the two contributions (\ref{7-5}), (\ref{7-11}) and using (\ref{5-4}), we obtain for
the mixing matrix
\begin{equation}\label{7a-3}
B^{'(1)}_{jk}
= 2 d^{1/2}_{jk}
\left(
2 S_{j} + S_{\frac{j-k}{2}}  - 2 S_{j-k} -  S_{\frac{j+k}{2}}
\right)
\, .
\end{equation}
Taking into account Eq.\ (\ref{7a-4}) and $\gamma^{(0)}_j=2S_{j}$ we finally arrive at
\begin{equation}\label{7a-5}
B^{(1)}_{jk} = 2 d^{1/2}_{jk} \left(
S_{j} + S_{k} + S_{\frac{j-k}{2}}-  S_{\frac{j+k}{2}} -2 S_{j-k} \right)\,.
\end{equation}
This result constitutes one of the main points of our paper. Namely,
we have been able to determine the first correction to the mixing
matrix of a particular type of twist-two operators in the $\cN=4$
SYM by just performing an order $g^1$ perturbative calculation. The
standard quantum field theory approach would require going to order
$g^4$ to obtain the same result. This clearly demonstrates the power
of conformal invariance applied to higher-loop calculations.

%%%%%%%%%%%%%%%%%%%%%%%%%%%%%%%%%%%%%%%%%%%%%%%%%%%%%%%%%%%%%%%%%%%%%
%            Figure
%%%%%%%%%%%%%%%%%%%%%%%%%%%%%%%%%%%%%%%%%%%%%%%%%%%%%%%%%%%%%%%%%%%%%
%%%%%%%%%%%%%%%%%%%%%%%%%%%%%%%%%%%%%%%%%%%%%%%%%%%%%%%%%%%%%%%%%%%%%

%%%%%%%%%%%%%%%%%%%%%%%%%%%%%%%%%%%%%%%%%%%%%%%%%%%%%%%%%%%%%%%%%%%%%
\section{Origin of the mixing matrix}
\label{Sect-OriMix}
%%%%%%%%%%%%%%%%%%%%%%%%%%%%%%%%%%%%%%%%%%%%%%%%%%%%%%%%%%%%%%%%%%%%%

As we established in the preceding sections, even in a theory with a vanishing $\beta-$function,
conformal symmetry is broken in the $\overline{\rm MS}$ scheme and leads to a mixing of conformal
composite operators under scale transformations. To analyze this mechanism in greater detail, we
employ conformal Ward identities \cite{Mue94,BelMue98c}. They can be simply derived form the
reparameterization invariance of the generating functional, given as a path integral. To derive
the true Ward identities it is crucial that the action contains a regulator, which in our case is
done by changing the integration volume, i.e., $d^4x \to d^Dx$ and replacing the coupling $g \to
\mu^{(D-4)/2} g$. Performing an infinitesimal field transformation $\Phi(x)\to \Phi^\prime(x) =
\Phi(x)+\delta\Phi(x)$ yields for a generic Green function of a (composite) operator $O(\Phi)$ to
the Ward identities,
\begin{equation}
\label{Gen-WI}
\langle
O(\Phi)  \delta\left(\Phi(x_1)\cdots \Phi(x_n)\right)
\rangle
=
- \langle
\left(\delta O(\Phi)\right) \Phi(x_1)\cdots \Phi(x_n) \rangle
- \langle O(\Phi) \left(\delta i S(\Phi)\right) \Phi(x_1)\cdots \Phi(x_n) \rangle,
\end{equation}
in the regularized theory. In a second step, we will perform the renormalization procedure, which
yields anomalous terms that emerge from contact terms in the product of the symmetry variation of
the regularized action functional and the operator insertion, i.e., $ O(\Phi) \left(\delta i S(\Phi)
\right) $. In particular when the conformal variation is involved, it generates the anomalous
dimensions and the mixing matrix. However, the source of the latter symmetry breaking is a peculiar
feature of dimensional regularization. In a third step, we employ conformal constraints, arising from
the conformal algebra, to perform a scheme transformation that restores conformal symmetry such that
conformal operators obey autonomous renormalization group equations.

Let us first with the conformal variations of the dimensional regularized  ${\cal N}=4$ SYM action
(\ref{Def-Act-N4}). Dilatation and special conformal boost variations lead to the following anomalous
operator insertions:
\begin{eqnarray}
\label{Var-Act-D}
\delta^D S \!\!\!&=&\!\!\! \varepsilon \int d^D x \left\{
\sum_{k = 1}^3 \mathcal{O}_{A_k} (x) + \mathcal{O}_{B} -
\Omega_\phi (x) - \Omega_{\lambda\bar\lambda} (x) \right\}
\, , \\
\label{Var-Act-K}
\delta^K_\mu S \!\!\!&=&\!\!\! \varepsilon \int d^D x \, 2
x_\mu \left\{ \sum_{k = 1}^3 \mathcal{O}_{A_k} (x) +
\mathcal{O}_{B} - \Omega_\phi (x) - \Omega_{\lambda\bar\lambda}
(x) \right\} - 2 (D - 2) \int d^D x \, \mathcal{O}_{\mu, B} (x)
\, . \
\end{eqnarray}
Here the relevant gauge invariant operator insertions, of type-A in terminology of Ref.\ \cite{KluZub75},
are
\begin{eqnarray}
\mathcal{O}_{A_1} (x) &=& {\rm tr} F_{\mu\nu} (x) F^{\mu\nu} (x)
\, , \\
\mathcal{O}_{A_2} (x) &=& \frac{g^2}{4} {\rm tr} [\phi^{AB} (x),
\phi^{CD} (x)] [\bar\phi_{AB} (x), \bar\phi_{CD} (x)]
\, , \\
\mathcal{O}_{A_3} (x) &=& \sqrt{2} g \, {\rm tr} \left\{
\bar\lambda_{\dot\alpha A} (x) [\phi^{AB} (x),
\bar\lambda^{\dot\alpha}_B (x)] - \lambda^{\alpha A} (x)
[\bar\phi_{AB} (x), \lambda_\alpha^B (x)] \right\} \, .
\end{eqnarray}
For the composite operator (\ref{3-12}) considered here, only the first two operator insertions
$\mathcal{O}_{A_i}$ can potentially contribute to our leading order analyses of conformal anomalies.
The type-B operators are BRST exact variations, and are irrelevant for the present study. Finally
$\Omega$ are equations of motion operator, e.g.,
\begin{eqnarray}
\Omega_\phi (x) = \frac{\delta S}{\delta \phi^{AB}} \phi^{AB} (x)
\, .
\end{eqnarray}

In the $\overline{\rm MS}$ scheme, the scale and special conformal Ward-identities for the Green
functions with the conformal operator $\hat{O}_j= \hat{O}_j(x=0)$ insertion, see Eq.~(\ref{3-12}),
can be found by plugging in the variations (\ref{Var-Act-D}) and (\ref{Var-Act-K}) into the generic
Ward identity (\ref{Gen-WI}),
\begin{eqnarray}
\label{Def-Reg-ConWI-D} \langle  [\hat{O}_j] \delta^D {\cal X}
\rangle &\!\!\! = \!\!\! &- \langle (\delta^D  [\hat{O}_j]) {\cal
X} \rangle - \langle  [\hat{O}_j]  (\delta^D i S) {\cal X} \rangle
\\
\label{Def-Reg-ConWI-K} \langle  [\hat{O}_j] \delta^K_- {\cal X}
\rangle &\!\!\! = \!\!\!  &- \langle (\delta^K_-  [\hat{O}_j])
{\cal X} \rangle - \langle  [\hat{O}_j] (\delta^K_- i S) {\cal X}
\rangle \, .
\end{eqnarray}
Without loss of generality, we specify the field monomial of elementary fields to the two scalar
fields of the theory ${\cal X} =\phi^{AB} (x_1)\phi^{CD} (x_2)$. By definition, the left-hand
side of these equations are finite, however, the separate terms on the right-hand side contain
anomalous contributions or even divergencies. To obtain the renormalized Ward identities we will
now renormalize the operator product of composite insertion and conformal variation of the action
and finally remove the regularization by sending $\varepsilon$ to zero.

In the case of the dilatation Ward identity (\ref{Def-Reg-ConWI-D}), the variation in the first
term on the right-hand side leads to the canonical dimension of the operator, while the second
term in the right-hand side of Eq.\ (\ref{Def-Reg-ConWI-D}) is ill-defined since it involves an
operator product at coincident space-time points,
\begin{eqnarray}
[\hat{O}_j]  (\delta^D S)
=
\varepsilon \int d^D x \, \mathcal{O}_{A_1} (x) [\hat{O}_j]
-
\int d^D x \, \Omega_\phi (x) [\hat{O}_j]
+ \cdots
\, .
\end{eqnarray}
Hence a subtractive renormalization procedure is required, where the divergence is a
$1/\varepsilon-$pole and its residue is nothing but the anomalous dimension of the operator.
The pole will be cancelled by the $\varepsilon-$term, appearing in the variation of the action
(\ref{Var-Act-D}). A rigorous treatment can be done by means of differential operator vertex
insertions \cite{Col84}. As expected, it can be demonstrated in a straightforward manner that
finally the Ward-identity (\ref{Def-Reg-ConWI-D}) turns into the renormalization group equation
for Green function with renormalized composite operator insertions,
\begin{eqnarray}
\label{Res-RGE}
\left[\mu\frac{\partial}{\partial\mu} + 2 \gamma_\phi(g^2)\right]
\langle  [\hat{O}_j] {\cal X} \rangle = - \sum_{k=0}^j
\gamma_{jk}(g^2) \langle  [\hat{O}_{jk}] {\cal X}\rangle\,.
\end{eqnarray}
Note that a $\beta-$function proportional term is absent in our conformal theory, nevertheless,
the trace anomaly in the regularized theory turns into the anomalous dimensions and, as we will
now see, it is also responsible for the mixing of operators.

%%%%%%%%%%%%%%%%%%%%%%%%%%%%%%%%%%%%%%%%%%%%%%%%%%%%%%%%%%%%%%%%%%%%%
%            Figure
%%%%%%%%%%%%%%%%%%%%%%%%%%%%%%%%%%%%%%%%%%%%%%%%%%%%%%%%%%%%%%%%%%%%%
\begin{figure}[t]
\begin{center}
\mbox{
\begin{picture}(0,135)(205,0)
\put(0,0){\insertfig{14.5}{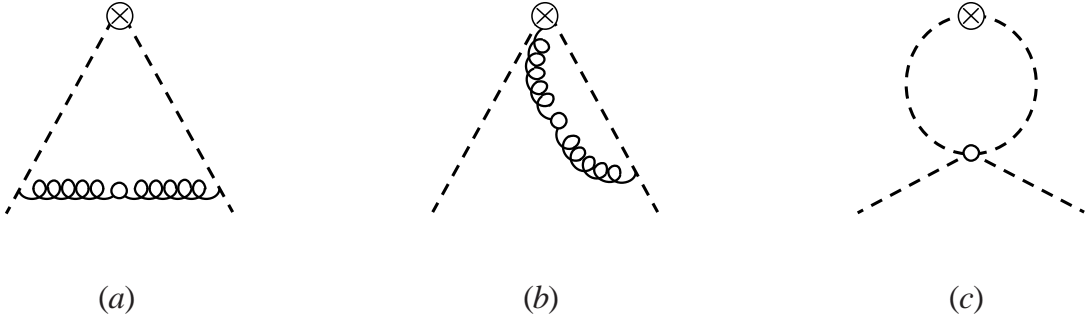}}
\end{picture}
}
\end{center}
\caption{ \label{FigRenOP} The symbol $\otimes$ stands for the
conformal operator and $\circ$ stands for the operator insertions
$\mathcal{O}_{A_i}$.}
\end{figure}%
%%%%%%%%%%%%%%%%%%%%%%%%%%%%%%%%%%%%%%%%%%%%%%%%%%%%%%%%%%%%%%%%%%%%%

The renormalization  procedure for the right-hand side of the special conformal
Ward identity (\ref{Def-Reg-ConWI-K}) has two peculiarities. First, the variation
of the renormalized operator insertion, i.e., $\delta^K_- [\hat{O}_j]$ leads to an
infinite expression which has to be cancelled against a singularity arising
from the renormalization of the operator product
\begin{eqnarray}
[\hat{O}_j]  (\delta^K_- S)
=
\varepsilon \int d^D x \, 2 x_- \mathcal{O}_{A_1} (x) [\hat{O}_j]
 -
\int d^D x \, 2 x_- \Omega_\phi (x) [\hat{O}_j] + \cdots.
\end{eqnarray}
Second, the subtractive renormalization of the latter is not automatically fixed by
the one appearing in the dilatation variation. The renormalized Ward identity can be
written as
\begin{eqnarray}
\label{WI-ConBos} \langle [\hat{O}_{jl}] \delta^K_- {\cal X}\rangle
=
- i \sum_{k = 0}^j \left[ a (l) + \gamma^c (l; g^2)
\right]_{jk} \langle  [\hat{O}_{k,l-1}] {\cal X} \rangle + \dots
\, ,
\end{eqnarray}
where the ellipsis contain BRST variations, the scaling dimensions in the conformal
variation  $\delta^K_-$ is now given by $1 + \gamma_\phi$, $a_{jk}(l) = 2 (l-k) (l+k+1)
\delta_{jk}$. Again in a conformal theory a $\beta$ proportional term is absent. The
so-called special conformal anomaly
\begin{eqnarray}
\label{Def-SpeConAno}
\gamma^c (l; g^2) = \frac{g^2}{(2\pi)^2} \left(Z^{[1] -}_{A_1} - 2 \gamma_\phi \, b
+ 2 [Z^{[1]}_{A_1}, b]\right) + O(g^2)\,,
\end{eqnarray}
where $b_{jk} = 2(j-k)(j+k+1) d^{1/2}_{jk}$, see Eq.\ (\ref{A2-2}), is expressed by two
subtractive renormalization constants $Z^{[1]}_{A_1} = \delta_{jk} (\gamma^{(0)}_j - 2
\gamma^{(0)}_\phi)/2$ and  $ Z^{[1] -}_{A_1}$. While the former is nothing else as the
residue of the renormalization matrix and is expressed in terms of the anomalous dimensions,
cf.\ Eq.\ (\ref{4-8}), the latter one is defined as
\begin{eqnarray}
\int d^D x \, 2 x_- \mathcal{O}_{A_1} (x) [\hat{O}_{jl}] =
\left[\int d^D x \, 2 x_- \mathcal{O}_{A_1} (x)
\hat{O}_{jl}\right]  +  \sum_{k=0}^j Z^-_{A_1, jk}
[\hat{O}_{k,l-1}]   + \cdots.
\end{eqnarray}
and is straightforwardly evaluated to lowest order  from the Feynman diagrams shown in Fig.\
\ref{FigRenOP}. The result is
\begin{eqnarray}
\label{Res-ZmA1}
Z^{-}_{A_1} = -\frac{g^2}{(2\pi)^2} \frac{1}{\varepsilon} \left(2 Z^{[1]}_{A_1}  b -
w \right)+ { O}(g^4)\,,
\end{eqnarray}
where the matrix elements of the $w$-matrix are given for $j>k$ by
\begin{equation}
\label{Res-w-Mat}
w_{jk}
=  2 (2 k+1) \left[1+(-1)^{j-k}\right]
\left(S_j +S_{\frac{j-k}{2}} -S_{\frac{j+k}{2}}- 2 S_{j-k}\right)\,.
\end{equation}
Combining Eqs.\ (\ref{Def-SpeConAno}), (\ref{Res-ZmA1}), and (\ref{Res-w-Mat}), we find that the
special conformal anomaly,
\begin{eqnarray}
\label{Res-SpeConAno}
\gamma^c_{j k}(j; g^2) = - 2(j-k) (j+k+1) \frac{g^2}{(2\pi)^2}  B^{(1)}_{jk} + O(g^4)\,,
\end{eqnarray}
is expressed by means of the mixing matrix $B^{(1)}_{jk}$, found earlier in Eq.\ (\ref{7a-5}).

The anomalous dimensions and the so-called special conformal anomaly are not entirely independent
quantities. In the case of a conformal field theory it is easy to derive a constraint between
them by acting with the differential operator $\mu d/d \mu$ on the conformal boost Ward identity
(\ref{WI-ConBos}) and with the generator of the conformal boost, --- a differential operator acting
on the arguments of elementary fields, --- on the renormalization group equation (\ref{Res-RGE}).
Subtracting both equations yields zero on the left-hand side while the right-hand side is the
desired constraint between the conformal anomalies, which we write as
\begin{eqnarray}
\label{Con-1}
2(j-k) (j+k+1) \gamma_{jk}(g) = \sum_{m=k}^j \left[\gamma_{jm}(g)
\gamma^c_{mk}(j; g) -\gamma^c_{jm}(j; g) \gamma_{mk}(g)\right].
\end{eqnarray}
The result can be understood as a consequence of the conformal commutator $[D,K_-] = i K_-$ applied
to the Green function with conformal operator insertion. We note that the additional $l-$dependence
in $\gamma^c_{jk}(l; g)$ is governed by another conformal constraint that arises from the commutator
relation $[K_-,P_+] = -2 i (D+M_{-+})$ and reads
\begin{eqnarray}
\label{Con-2} \gamma^c(l+1; g) = \gamma^c(l; g) -2 \gamma(g)\,.
\end{eqnarray}
The above conformal constraints guarantee that there exist a scheme in which the covariant behavior of
conformal operator under conformal transformations is ensured \cite{Mul97}. Such a scheme is obtained
by a finite renormalization group transformation (\ref{Dia-AD}) that diagonalizes the anomalous dimension
matrix. Utilizing the constraints (\ref{Con-1}) and (\ref{Con-2}) one finds the $B$-matrix in terms of
the special conformal anomaly:
\begin{eqnarray}
B = \frac{1}{1+J \gamma^c} = 1 - J\gamma^c + J(\gamma^c J
\gamma^c) -\cdots\,,\quad \mbox{where}\quad (J\gamma^c)_{jk}  =
\frac{\gamma^c_{jk}}{2(j-k) (j+k+1) }\,.
\end{eqnarray}
Needless to say that the mixing matrix to order $g^2$ follows by inserting the expression
(\ref{Res-SpeConAno}) for the special conformal anomaly into this equation and coincide with
the result (\ref{7a-5}), obtained in Sect.\ \ref{Sec-N4SUSY}. Rotating now the conformal
operator via Eq.\ (\ref{4-6a}), the conformal boost Ward identities (\ref{WI-ConBos}) turns
into \cite{Mul97}:
\begin{eqnarray}
\label{WI-ConBos-1} \langle \hat{\mathbb{O}}_{jl} \delta^K_- {\cal X}
\rangle = - i\, 2(j-l)\left(j+l+1+\gamma_j(g^2)\right) \langle
[\hat{\mathbb{O}}_{j,l-1}] {\cal X} \rangle + \dots.
\end{eqnarray}
Here the ellipsis stands for Green functions with BRST-exact operator insertions $\mathcal{O}_B$,
which do not contribute in gauge invariant quantities. As we see, the conformal operators
$\mathbb{O}_{jl}$ transform covariantly under conformal boost and in particular the lowest state
in the module $\mathbb{O}_{j}$ is invariant under conformal boost.

%%%%%%%%%%%%%%%%%%%%%%%%%%%%%%%%%%%%%%%%%%%%%%%%%%%%%%%%%%%%%%%%%%%%%
\section{Conclusions}
%%%%%%%%%%%%%%%%%%%%%%%%%%%%%%%%%%%%%%%%%%%%%%%%%%%%%%%%%%%%%%%%%%%%%

As we demonstrated in this work, in a conformal field theory there
exists a special renormalization scheme in which renormalized
conformal operators possess diagonal two-point correlation functions
(\ref{3-1}), whose fall-off with distance is determined merely by their scale
dimensions. We proposed to use these correlators for evaluation of
anomalous dimensions of conformal operators at higher orders making
use of the so-called Anselmi's trick. The main advantage of this
formalism for multi-loop calculations of leading twist anomalous
dimensions, compared to diagrammatic evaluation of operator matrix
elements, arises from simpler topologies of contributing Feynman
graphs taking the form of bubble diagrams. Along this line of
reasoning, the knowledge of the four loop diagrams, appearing in the
descendants, would permit a three-loop calculation of anomalous
dimensions.

We finally remark that in a non-conformal theory, e.g., QCD, the
covariant behavior of conformal operators is spoiled by a term
proportional to the $\beta-$function. Hence, the form of the
correlation functions will change too. However, setting by hand the
$\beta-$function to zero at a given order of perturbation theory,
the conformal prescription still applies and can be used for highly
non-trivial predictions. We can even argue that in certain cases the
trace anomaly can be entirely incorporated into the conformal
predictions. For instance, we can introduce a scheme in which the
conformal operators of leading twist are multiplicatively
renormalizable. Then the form of the two-point correlation function
in the full theory is entirely fixed by the renormalization group
equation:
\begin{equation}
\langle \hat{\mathbb{O}}_j(x_1)\, \hat{\mathbb{O}}_k(x_2)
\rangle_{(\mu^2)}
=
\delta_{jk} C_j \left( \bar{g} ( 1/{\scriptscriptstyle \sqrt{x^2_{12}}};g) \right)
\
\hat{I}^{j}\  (x_{12}^2)^{-d_j} \exp\left\{
\int_{1/\sqrt{x_{12}^2}}^{\mu} \frac{d\mu^\prime}{\mu^\prime}
\gamma_j(\bar{g}(\mu^\prime;g)) \right\},
\end{equation}
where the running coupling  satisfies the initial condition $\bar{g}(\mu;g)= g$.

Another potential application of the method is the evaluation of the
mixing matrix in full QCD to two-loop order accuracy, including the
$\beta-$function (see, e.g., Ref.\ \cite{BelMue98c}). This allows one to
restore the full anomalous dimension matrix and even the non-forward
evolution kernels in the $\overline{\rm MS}$ scheme to three-loop
level. This piece of information is required in the application of
perturbative QCD to exclusive processes and would allow for a
complete next-to-next-to-leading analysis in the $\overline{\rm MS}$
scheme.

\vspace{5mm}

We benefited from enlightening discussions with B.\ Eden, J.M.\
Drummond, G.P.\ Korchemsky and I.T.\ Todorov. This work was
supported by the U.S. National Science Foundation under grant no.
PHY-0456520 (A.B.) and the French Agence Nationale de la Recherche
under contract ANR-06-BLAN-0142 (E.S.). Three of us (A.B., J.H.\ and
E.S.) would like to thank LPT (Orsay) and (A.B.\ and D.M.) are
grateful to LAPTH (Annecy) for the warm hospitality extended to them
at different stages of the work. J.H.\ acknowledges the warm
hospitality extended to him by the Theory Group of the Dipartimento
di Fisica, Universit\`{a} di Roma ``Tor Vergata''.

\appendix
%%%%%%%%%%%%%%%%%%%%%%%%%%%%%%%%%%%%%%%%%%%%%%%%%%%%%%%%%%%%%%%%%%%%%
\section{Appendix: Calculation of descendants}
\label{App-des}
%%%%%%%%%%%%%%%%%%%%%%%%%%%%%%%%%%%%%%%%%%%%%%%%%%%%%%%%%%%%%%%%%%%%%

We calculate the divergence (\ref{2-16}) of the twist-two operator
(\ref{3-12}) on the classical level, using the equations of motion.
The terms in the descendant $K$ have two origins: the commutation
relation of covariant derivatives and the use of the Klein-Gordon
equation derived from the action (\ref{Def-Act-N4}). The final
result can be written as
\begin{eqnarray}\label{A1-6}
\hat{K}_{j-1} &=& A_{j}(\hat{\pa}_{a},\hat{\pa}_{b},\hat{\pa}_{c})
\; \mathrm{tr}\; \left( \{ \lambda^{\alpha}_{3}(x_a),
\lambda_{\alpha 4}(x_c)\} \phi^{12}(x_b) + \{
\bar{\lambda}_{\dot{\alpha}}^{1}(x_a), \bar{\lambda}^{\dot{\alpha}
2}(x_c)\}
\phi^{12}(x_b) \right)  \nonumber \\
&+&
B_{j}(\hat{\pa}_{a},\hat{\pa}_{b},\hat{\pa}_{c})
\;
\mathrm{tr}\;
{}[
\partial^{\mu} \phi^{12}(x_a), \phi^{12}(x_b) ] F^{\mu\nu}(x_c) z_{\nu}
+ O(g)\,,
\end{eqnarray}
where $A_{j}$ and $B_{j}$ are homogeneous polynomials of degree $j-1$ and $j-2$, respectively.
Here we have dropped $O(g)$ terms, which are quadratic in $\phi$ or contain two $F^{\mu\nu}$'s,
since they are irrelevant for the considered perturbative order. In
terms of new variables $u=a+b+c$, $s=(a+c-b)/u$ and $t=(a-c-b)/u$
they can be written as
\begin{equation}\label{A1-7}
A(a,b,c) = 2\sqrt{2} u^{j-1} a_{j}(s)\,,\qquad  B_{j}(a,b,c) = -i
8 u^{j-2} b_{j}(s,t)
\end{equation}
with
\begin{eqnarray}\label{A1-8} b_{j}(s,t) &=& \frac{1}{s-t} \left[
P^{'}_{j}(s) + (s-1) P^{''}_{j}(s)\right]+ \frac{1}{(s-t)^2}
\left[(1-s) P^{'}_{j}(s) - (1-t) P^{'}_{j}(t)\right] \\
a_{j}(s) &=& P_{j}^{'}(s) + (s-1) P_{j}^{''}(s)\,,
\end{eqnarray}
and $P_{j}(x)=C_{j}^{1/2}(x)$. Note there is no singularity in $b_{j}(s,t)$ for $s=t$, which can
be seen by Taylor expanding around $s=t$.

\end{document}